\documentclass[12pt,preprint]{aastex}

\shorttitle{What can the redshift tell us?}

\shortauthors{Xu}

\input psfig.sty

\begin{document}

\title{What can the redshift observed in EXO 0748-676 tell us?}

\author{Ren-Xin Xu}

\affil{Astronomy Department, School of Physics, Peking University,
Beijing 100871, China}

\email{rxxu@bac.pku.edu.cn}

\begin{abstract}

The mass-radius relations for bare and crusted strange stars are
calculated with the bag model. Comparing these relations with the
observed one derived from the redshift, we address that the
conclusion,  that EXO 0748-676 can not be a strange star, is
incorrect.
Various strange star models can show that EXO 0748-676 could have
a mass of $(1.3\sim 1.7) M_\odot$ and a radius of $(8.4\sim 11.4)$
km.
It is proposed that part of nascent strange stars could be bare
and have mass $\sim 0.1M_\odot$, whereas their masses increase
during a long accretion history.

\end{abstract}


\section{Introduction}

On one hand, identifying strange stars is among the most important
problems in the new millennium astrophysics.
Strange stars are hypothetical compact objects that consists of
roughly equal numbers of deconfined up, down and strange quarks,
to affirm or negate the existence of which should have great
implications in the study of the elemental strong interaction
(see, e.g., Xu 2002a, for a review).
Some compact objects, previously known as neutron stars, may
actually be strange stars. Whatever they are, the most essential
and important thing is to find {\em clear} observational
signatures of strange stars.

However, on the other hand, {\em how to identify a neutron star?}
If a neutron star is found with certain, we are very probably to
obtain a negative conclusion of strange stars eventually.
A neutron star, as its name implies, is made mainly of neutrons,
the outmost part of which is an atmosphere of normal ions.
Recently, it is a central goal and a real competition among the
observers to find line emission from the atmosphere, since the
stellar mass $M$ and radius $R$ may be derived by obtaining its
gravitational redshift (as $M/R$) and the pressure broadening (as
$M/R^2$) of the lines.
Still almost no line is observed (Xu 2002b) except for those two
sources 1E 1207.4-5209 and SGR 1806-20 (Xu et al. 2002).

It is possible that strong magnetic field, $\ga 10^{12}$ G, around
a neutron star may greatly modify the thermal spectrum, making
lines difficult to identify (Pons et al. 2002).
A recent research of studying EXO0 748-676, a compact star that
has much weaker field ($\sim 10^8$ G), shows significant
absorption lines, the Fe XXVI and XXV $n=2\rightarrow 3$ and the O
VIII $n=1\rightarrow 2$, in the spectra of 28 bursts\footnote{%
EX O0748-676 is an X-ray burster; the mechanism responsible for
the bursts is attributed to the nuclear fusion on the surface due
to enough accretion from its low mass company.
}. %
All of these lines are redshifted, with an unique value of 0.35
(Cottam et al. 2002).
The authors concluded then, according to the redshift, that their
results are expected for neutron star models, but do not agree
with a strange star model based on the equation of state proposed
by Dey et al. (1998) if the mass of EXO 0748-676 is greater than
$1.1M_\odot$.

It will be addressed in this paper that strange star models for
EXO 0748-676 can not be excluded. According to a simplified
version of the MIT bag model, we calculate the mass-radius
relations of strange stars and compare them with observations, and
find that it is very reasonable to assume that EXO 0748-676 is a
strange star.

\section{The mass-radius relations}

In the outer vacuum of a spherically symmetric object with mass
$M$, due to the general relativity effects, a photon with
wavelength $\lambda_0$, radiated at radius $r$, should be
red-shifted. The received wavelength at infinity is
$\lambda=\lambda_0/\chi$, where $\chi=\sqrt{1-2GM/(c^2r)}$ is
called as the redshift factor ($c$ is the speed of light).
The redshift is defined as $z=(\lambda-\lambda_0)/\lambda_0$, and
one thus have a relation,
\begin{equation}
M_1=3.37[1-(1+z)^{-2}]r_6=1.52r_6\;\; ({\rm for}\; z=0.35),
\label{m-r}
\end{equation}
where $M_1=M/M_\odot$, and $r_6=r/(10^6$cm). The second equation
above is for $z=0.35$ observed in EXO 0748-676.
If photons are emitted at stellar surface, $r=R$ ($R$ is the
stellar radius), Eq.(\ref{m-r}) is actually the mass-radius
relation of a central object, which is a beeline the M-R diagram.

Can strange star models satisfy this relation?
Actually strange star structures with crusts were discussed
previously (e.g., Kettner et al. 1995, Huang \& Lu 1998, Madsen
1999).
We need an equation of state of strange matter in order to model a
strange star.
In a simplified version of the bag model, assuming quarks are
massless, we then have quark pressure $P_{\rm q}=\rho_{\rm q}/3$
($\rho_{\rm q}$ is the quark energy density); the total energy
density is $\rho=\rho_{\rm q}+B$ but the total pressure is
$P=P_{\rm q}-B$. One therefore have the equation of state for
strange matter (Alcock et al. 1986),
\begin{equation}
P=(\rho-4B)/3.
\label{eos}
\end{equation}
Although the equation is simple, the crucial parameter, the so
called bag constant $B$, can hardly be given.
Nevertheless, the preferred value for $B$ is in the range of $60
{\rm MeV/fm^3}\la B \la 110 {\rm MeV/fm^3}$, according to the
studies of the hadronic spectrum, the hadronic structure
functions, and the comparison of bag model with lattice quantum
chromodynamics (Drago 2000).
We will thus calculate with $B=60 {\rm MeV/fm^3}$ and $110 {\rm
MeV/fm^3}$ for indication.

In an accreting binary system, it is possible that a strange star
could be crusted. The mass and the height of the crust can be
calculated by 1, the density, $\rho_{\rm b}$, at the base of the
crust, and 2, the gravitational equilibrium with suitable equation
of state of matter in the crust.
The density $\rho_{\rm b}$ can not be higher than the density of
neutron drip $\rho_{\rm d}=4\times 10^{11}$ g/cm$^3$, because free
neutrons, which will not fell Coulomb force and should melt in the
strange quark matter, appear at higher density. The crust height
computed with $\rho_{\rm b}=\rho_{\rm d}$ is an upper limit, and
the stellar radius (or mass) is thus between that of bare strange
star and that of crusted strange star with $\rho_{\rm b}=\rho_{\rm
d}$.
The standard equation of state of cold, fully catalyzed matter
below neuron drip is given by Baym, Pethick \& Sutherland (1971),
which is called as BPS equation of state.

The mass-radius relations of strange stars are calculated with
Eq.(\ref{eos}) for the strange quark matter core and with BPS
equation of state for the crust (we choose $\rho_{\rm b}=\rho_{\rm
d}$), which are shown in Fig.1.
It is found from Fig.1 that the redshift of $z=0.35$ can easily be
consistent with strange star models, at least in the regime of
strange quark matter described by MIT bag model.
For $B=110$ MeV/fm$^3$, EXO 0748-676 could have a mass of $\sim
1.3 M_\odot$ and a radius of $\sim 8.4$ km, whereas for $B=60$
MeV/fm$^3$, it could have $M \sim 1.7 M_\odot$ and $R\sim 11.4$
km.
Therefore it is possible that the mass of EXO 0748-676, if being a
strange star, can be larger than $1.1 M_\odot$, which is certainly
comfortable from astrophysical arguments.

\section{Conclusions and Discussions}

Using MIT bag model and BPS equation of state, we calculate the
mass-radius relations for both bare and crusted strange stars, and
compare them with the observed one derived from the redshift. It
is found that we still can not rule out strange star models for
the X-ray burster EXO 0748-676.
According to the calculations presented, EXO 0748-676 could have a
mass of $(1.3\sim 1.7) M_\odot$ and a radius of $(8.4\sim 11.4)$
km.

There might be some observational indications of very-low-mass
compact stars, with mass $\sim 0.1M_\odot$, which could be
representative of strange stars. 1. From the 500 Ksec {\em
Chanrda} record of thermal spectrum of the nearest and brightest
isolated neutron star RX J1856-3754, fitted with a single
temperature Plank spectrum, one can deduce a radius
$R_\infty=3.8\sim 8.2$ km (Pons et al 2002, Drake et al. 2002),
although these results were soon criticized by Walter \& Lattimer
(2002).
It is worth noting that the apparent radius $R_\infty$, i.e., the
values observed at infinity, is not that calculated in
Eq.(\ref{m-r}). For a compact star with mass $M$ and radius $R$,
the relation is (e.g., Haensel 2001)
\begin{equation}
R_\infty = R/\sqrt{1-(R_{\rm s}/R)},
\end{equation}
where $R_{\rm s}=2GM/c^2$ is the Schwartzschild radius.
According to Haensel's (2001) results for bare strange stars [RX
J1856-3754 could be a bare strange star because of its featureless
spectrum, see Xu (2002b) for details], RX J1856-3754 may therefore
have a mass of $(0.06\sim 0.4) M_\odot$.
2. Based on the study of radio pulse beam and its polarizations,
it is suggested that the fastest rotating pulsar, PSR 1937+21,
could be a strange star, with mass $M<0.2M_\odot$ and radius $R<1$
km (Xu et al. 1999).
If this very-low-mass strange star was born with a period $P_0\la
1.56$ ms, in order to prevent it from developing the rotation-mode
instability (e.g., Madsen 1998), the star can have a minimum
period $P_{\rm min}\sim 0.1$ ms (or angular frequency $\Omega\sim
6\times 10^6$/s) in case of $M=0.2M_\odot$, $R=1$ km, and the
temperature of a nascent strange $T=10^9$ K.
Therefore PSR 1937+21 may not have an accretion history since
$P_0\gg P_{\rm min}$.
3. Also the gravitational microlensing study reports events with a
mass $\sim 0.5M_\odot$ towards LMC (Alcock et al. 2000), and with
$M=0.13M_\odot$ and $M=0.25M_{\rm Jupiter}$ at the globular
cluster M33 (Sahu et al. 2001).

Combining the results of masses and radii of very-low-mass compact
stars and of EXO 0748-676, we may conjecture, that part of bare
strange stars may have low masses, $\sim 0.1M_\odot$, while
strange stars with much high accretion rate history could have
high masses, $\sim 1.5M_\odot$.
A natural explanation for this is that long historical accretion
increases the masses of strange stars.
Therefore, it is possible that some of the newborn strange stars,
which should be bare (Xu 2002b), could have much small masses and
radii.
What is the critical physical reason which determines the initial
mass of a strange star? This is one of the interesting topics in
the study of supernova explosion.

The strange star model can not be excluded even for the equation
of state of Dey et al. (1998). No solid evidence to show that EXO
0748-676 has a mass $\ga 1.1 M_\odot$. The star can accrete much
to a mass $\sim 1 M_\odot$ during a long accretion history if its
initial mass is very low.

\vspace{0.2cm} \noindent {\it Acknowledgments}:
This work is supported by National Nature Sciences Foundation of
China (10273001) and the Special Funds for Major State Basic
Research Projects of China (G2000077602).
I would like to thank Mr. Yi Liu for technique help in preparing
the figure.

\begin{figure}

\centerline{\psfig{figure=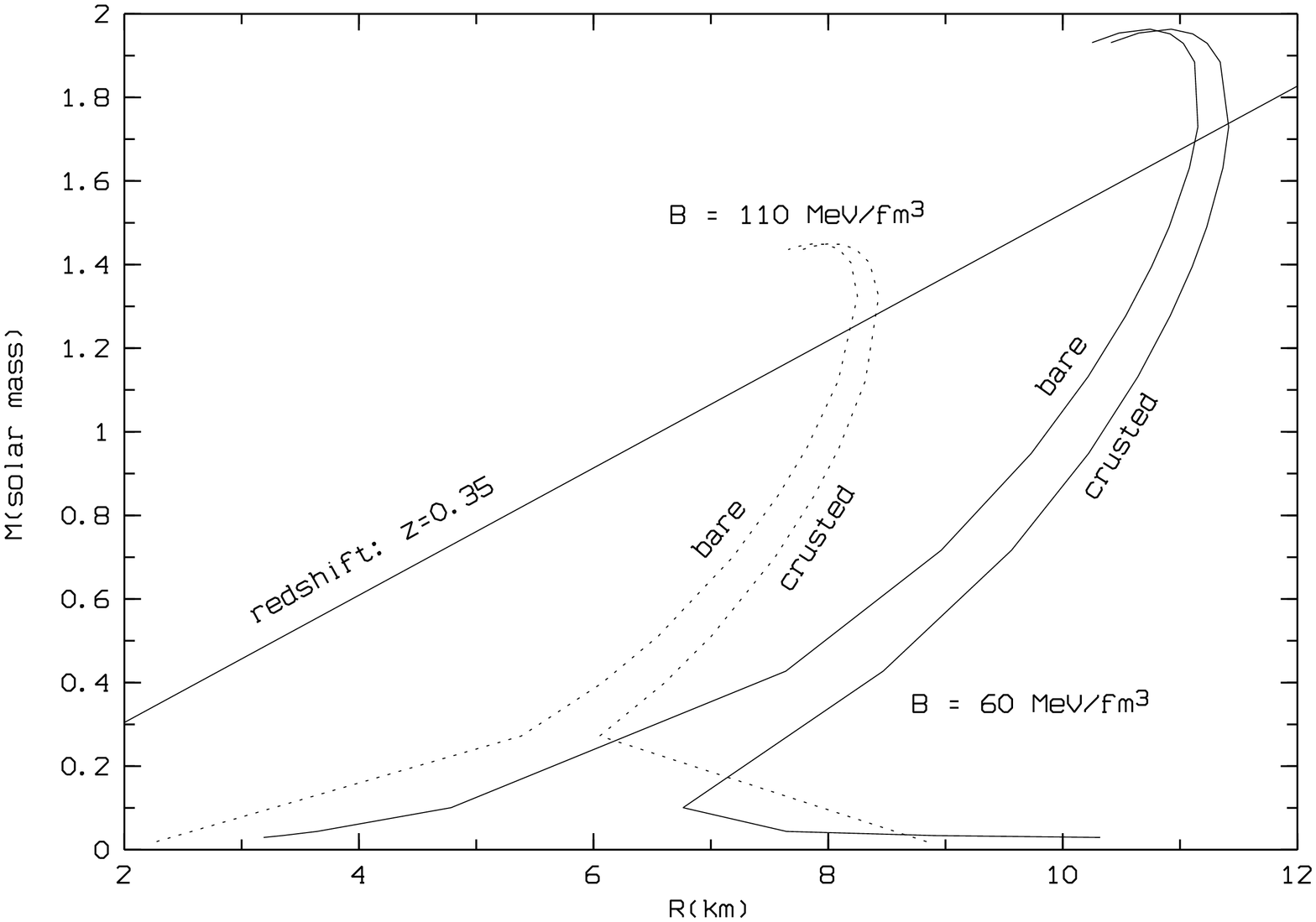,height=100mm,width=120mm,angle=270}}
\caption{The mass-radius relations for strange stars, both bare
and crusted, based upon a simplified version of MIT bag models.
Solid and dotted lines are for bag constant $B=60$ MeV/fm$^3$ and
110 MeV/fm$^3$, respectively. The mass-radius relation derived
from the redshift $z=0.35$ is also shown.
\label{Fig1}}
\end{figure}

\end{document}